\documentclass[12pt]{iopart}

\usepackage{citesort}

\usepackage{color}
\usepackage{graphicx}
\usepackage{psfrag}

\psfrag{Ccal}[B1][B1][1]{${\mathcal{C}}/R$}
\psfrag{Ccal1}[B1][B1][1]{${\mathcal{C}}$}
\psfrag{Vcal}[B1][B1][1]{${\mathcal{V}}$}
\psfrag{Ecal}[B1][B1][1]{${\mathcal{E}}$}

\begin{document}

\title[Simulations of the RPM at low temperature and density]{Computer 
simulations of the restricted primitive model at very low temperature and 
density}

\author{
Chantal Valeriani$^{1,2}$, 
Philip J. Camp$^{3}$, 
Jos W. Zwanikken$^{4,5}$,
Ren{\'e} van Roij$^{5}$ and
Marjolein Dijkstra$^{2}$
}

\address{1. SUPA, School of Physics and Astronomy, The University of 
Edinburgh, Mayfield Road, Edinburgh EH9 3JZ, United Kingdom.}

\address{2. Soft Condensed Matter, Debye Institute for NanoMaterials 
Science, Utrecht University, Princetonplein 5, 3584 CC Utrecht, The 
Netherlands.}

\address{3. School of Chemistry, The University of Edinburgh, West Mains 
Road, Edinburgh EH9 3JJ, United Kingdom.}

\address{4. Department of Material Science and Engineering, Northwestern 
University, 2220 Campus Drive, Evanston, Illinois 60208-3108, United 
States.}

\address{5. Institute for Theoretical Physics, Utrecht University, 
Leuvenlaan 4, 3584 CE Utrecht, The Netherlands.}

\ead{cvaleria@ph.ed.ac.uk}

\begin{abstract} The problem of successfully simulating ionic fluids at 
low temperature and low density states is well known in the simulation 
literature: using conventional methods, the system is not able to 
equilibrate rapidly due to the presence of strongly associated 
cation-anion pairs. In this manuscript we present a numerical method for 
speeding up computer simulations of the restricted primitive model (RPM) 
at low temperatures (around the critical temperature) and at very low 
densities (down to $10^{-10}\sigma^{-3}$, where $\sigma$ is the ion 
diameter). Experimentally, this regime corresponds to typical 
concentrations of electrolytes in nonaqueous solvents. As far as we are 
aware, this is the first time that the RPM has been equilibrated at such 
extremely low concentrations. More generally, this method could be used to 
equilibrate other systems that form aggregates at low concentrations. 
\end{abstract}


\section{Introduction}

Computer simulations have yielded invaluable insights on the properties of 
ionic fluids. The nature of fluid-fluid (`vapour-liquid') phase separation 
and the universality class of the associated critical point have attracted 
particular attention. In these studies, the restricted primitive model 
(RPM) has played a central role 
\cite{debye1923,larsen1974,stell1976,friedman1979,henderson1983,%
schreiber1987,pitzer1990,valleau1990,pitzer1992,stell1992,orkoulas1994,%
panagiotopoulos1994,caillol1994,stell1996,fisher1996,camp1999JCP,luijten2001}. 
The RPM is a simple representation of molten salts and ionic solutions. It 
consists of an equimolar binary mixture of positively and negatively 
charged hard spheres with charges $\pm q$ and equal diameters $\sigma$, 
immersed in a continuum with dielectric constant $\epsilon$. In terms of 
the model parameters, the reduced temperature is defined as 
$T^{*}=k_{B}TD\sigma/q^{2}$, where $k_{B}$ is the Boltzmann's constant, $T$ is 
the absolute temperature, $D=4\pi\epsilon\epsilon_{0}$, and $\epsilon_{0}$ 
is the vacuum dielectric permittivity; the reduced density is defined as 
$\rho^{*}=\rho\sigma^{3}$, where $\rho=N/V$ is the total number of ions 
per unit volume. The most recent high-precision Monte Carlo (MC) 
simulations locate the critical point at a critical temperature $T_{c}^{*} 
\simeq 0.05$ and a critical density $\rho_{c}^{*} \simeq 0.08$ 
\cite{caillol2002,luijten2002}; the critical point has been confirmed as 
belonging to the three-dimensional Ising universality class 
\cite{luijten2002}. Interestingly, the critical temperature is close to 
room-temperature conditions for sub-nanometre monovalent ions in oily 
solvents with $\epsilon \simeq 5$-$10$. However, the ion concentrations in 
these nonaqueous electrolyte solutions are often in the nM regime 
($\rho^{*} \sim 10^{-10}$) 
\cite{pnasleunissen2007,zwanikken2007,leunissen2007} which motivates the 
parameter choice of the present study.

It has been clear for a long time that physical clustering of the ions has 
an important effect in the vapour region \cite{friedman1979,friedman1981}, 
as strongly suggested by the Bjerrum theory \cite{bjerrum1926}. Analysing 
the features of this system in the low density--low temperature regime, 
Valleau \cite{valleau1980} and Gillan \cite{gillan1983} showed that the 
ionic fluid tends to form dimers, triplets, and higher order clusters, and 
that clustering has a crucial effect on the equilibrium properties of the 
RPM. Weis and Caillol \cite{caillol1995} and Bresme {\it et al.} 
\cite{bresme1995} characterised the cluster structures quantitatively at 
temperatures around $T_{c}$ and at densities around $\rho_c/5 $ and 
$\rho_c/3$, respectively. Later on, Camp and Patey identified different 
regimes of ion association well below $\rho_{c}$ \cite{camp1999PRE}: at 
low temperature the system apparently consists of only clusters; at 
intermediate temperature the system is predominantly associated, but with 
some free ions; and at high temperature the majority of ions are free. The 
RPM with {\em screened} Coulombic interactions can serve as a model for 
charged colloidal systems \cite{caballero2004,hynninen2006,fortini2006}. 
Caballero and coworkers studied such a model with an inverse screening 
length of $\kappa=6\sigma^{-1}$, mimicking the effect of added 
electrolytes present in the medium: the critical point was located at 
$T_{c}^{*} \simeq 0.17$ and $\rho_{c}^{*} \simeq 0.22$, and the familiar 
clustering phenomenon (of the colloids) in the dilute phase was observed.

The main obstacle to simulating ionic fluids successfully in the 
low-temperature regime, where coexistence occurs, is the strong 
association of ions at distances close to contact and the resulting 
extremely slow equilibration. Graham and Valleau \cite{valleau1990} 
pointed out that, when studying the low-temperature regime, conventional 
MC or molecular dynamics methods are not sufficient to equilibrate the 
system. Therefore, they first used a type of umbrella sampling named 
``temperature scaling Monte Carlo'' at several densities 
\cite{torrie1977}. Next, Valleau proposed ``density scaling Monte Carlo", 
a novel algorithm based on umbrella sampling over broad ranges of 
densities \cite{valleau1991-1}, and applied it to the RPM near the 
critical point. Orkoulas and Panagiotopoulos computed the vapour-liquid 
phase diagram \cite{orkoulas1994}, and to accelerate convergence, proposed 
ion-pair and cluster moves capable of grouping and moving clustered ions. 
The primary motivation in this work was the computation of the 
vapour-liquid phase diagram, and hence reduced densities of no less than 
$10^{-4}$ were considered.

In recent work, Allahyarov and co-workers \cite{lowen} have shown that, 
around the critical temperature, oppositely charged micro-ions tend to 
form `Bjerrum pairs', in which oppositely charged particles are closer 
than the Bjerrum length $\lambda_{B}=q^2/k_{B}TD$. The authors of 
reference \cite{lowen} found that at the lowest salt concentrations 
(around $10^{-9}~\mbox{mol}~\mbox{L}^{-1}$, corresponding to a reduced 
density of around $10^{-10}$ for ions with $\sigma = 5~\mbox{\AA}$), 
almost $90\%$ of the ions resided in pairs. In a later publication 
\cite{erratum}, the same authors used a different definition of a cluster 
(with a cut-off of $\lambda=3\sigma$), and narrowed their results down to 
a smaller concentration range: the new results became valid only for salt 
concentrations between $10^{-4}~\mbox{mol}~\mbox{L}^{-1}$ and 
$10^{-2}~\mbox{mol}~\mbox{L}^{-1}$ (or reduced densities between $10^{-5}$ 
and $10^{-3}$). The main problem found by the authors was equilibrating 
the system at extremely low densities by means of standard simulation 
techniques.

The aim of our work is to present a novel MC technique that achieves rapid 
equilibration of the RPM at low temperatures (around $T_{c}$) and very low 
reduced densities (from $10^{-3}$ down to $10^{-10}$). This method might 
also be applied to the equilibration of other systems that form aggregates 
at low concentrations.

\section{Simulations}

The interaction potential for the RPM is
\begin{equation}
\label{rpm}
U(r_{ij}) =
\left\{
\begin{array}{ll}
\infty   &  r_{ij} <   \sigma \\
 q_i q_j  / Dr &  r_{ij} \ge \sigma,
\end{array}
\right.
\end{equation}
where $q_i=\pm q$. The system is comprised of $N/2$ cations and $N/2$ 
anions in a cubic box of length $L$, with periodic boundary conditions 
(PBCs) applied. We use MC simulations of $N=256$ ions in the $NVT$ 
ensemble. The choice of such a relatively small number of particles is 
justified by the fact that, as the densities under study are so low, the 
simulation box is always large enough to exclude any significant 
finite-size effects due to the PBCs; with $N=256$ ions and $\rho^{*} 
\propto 10^{-10}$, the simulation box length is around $10^4 \sigma$. We 
checked for finite-size effects by running simulations at 
$\rho^{*}=10^{-4}$ with either 256 or 1000 particles, and making sure that 
the computed energy per particle was the same within statistical 
uncertainties. Moreover, the box lengths are large compared to the range 
of Debye-like screening, equal to the Debye length $\lambda_{D}/\sigma = 
\sqrt{ T^{*} / 4 \pi \rho^{*} }$. Table \ref{tavola1} shows the values of 
$L$ and $\lambda_{D}$ at the densities and temperatures considered in our 
work; the density range is $10^{-10} \leq \rho^{*} \leq 10^{-3}$ and the 
temperature range is $0.04 \leq T^{*} \leq 0.07$. In this paper we will 
concentrate on simulations at the lowest density and temperature; the full 
range of state points will be considered in a forthcoming publication.

\begin{table}[ht]
\caption{Box edge $L$, Debye screening length $\lambda_D$, and Ewald 
real-space screening parameter $\alpha$ for all of the simulated densities 
and temperatures ranging from $T^{*}=0.04$ to $T^{*}=0.07$, and with 
$N=256$. For each density, the smallest value of $\lambda_{D}$ corresponds 
to the lowest temperature, and the largest $\lambda_{D}$ to the highest 
temperature.\label{tavola1}}
\begin{center}
\begin{tabular}{cccc}\hline
$\rho^{*}$             & $L/\sigma$ & $\lambda_D/\sigma$  & $\alpha\sigma$ \\
\hline 
$1.73 \times 10^{-1}$  &    11.39   & $0.135$--$0.179$      & $0.49$ \\ 
$8.68  \times 10^{-2}$ &    14.34   & $0.191$--$0.253$      &  $0.39$ \\
$2.73  \times 10^{-3}$ &    45.45   & $1.080$--$1.428$      & $0.12$ \\ 
$1.10  \times 10^{-3}$ &    61.48   & $1.700$--$2.249$      & $0.091$ \\ 
$1.00  \times 10^{-4}$ &    136.13  & $5.642$--$7.460$      &  $0.041$ \\ 
$6.70  \times 10^{-6}$ &    336.75  & $21.790$--$28.812$    & $0.017$ \\ 
$2.29  \times 10^{-6}$ &    481.62  & $37.283$--$49.296$    & $0.012$ \\ 
$1.05  \times 10^{-6}$ &    624.44  & $55.059$--$72.800$    & $0.0090$ \\  
$9.48  \times 10^{-9}$ &   2999.78  &$579.457$--$766.167$   & $0.0019$ \\  
$9.03  \times 10^{-11}$ &  14154.79  & $5937.193$--$7850.244$ & $0.00040$ \\
\hline 
\end{tabular}
\end{center}
\end{table}

The long-range interactions were handled using the Ewald sum with tin-foil 
boundary conditions \cite{sanz,dlpoly,ewald1,ewald2}. For each density we 
carefully tuned the Ewald parameters $\alpha$, $r_c$ and $k_{\rm max}$, 
being the width of the Gaussian distribution characterising the screening 
term in real space, the real-space cut-off, and the reciprocal-space 
cut-off, respectively. $\alpha$ was chosen using the empirical rule 
$\alpha L = 5.6$ \cite{allen1987}, $r_c$ was set to $L/2$, and $k_{\rm 
max}$ such that the relative error in the reciprocal-space sum was of the 
order of $10^{-5}$ \cite{dlpoly}. The values of $\alpha$ are indicated in 
table \ref{tavola1}; $k_{\rm max}$ was always set to $10\times (2\pi/L)$. 
To test our code, we computed the energy per ion pair $u_p$ in a liquid at 
$T^{*}=0.042$ and $\rho^{*}=0.17$; we obtained $\langle \beta u_{p} 
\rangle=-1.26 \pm 0.01$, which is in perfect agreement with that computed 
for the same state point by Romero-Enrique {\it et al.} \cite{romero2002}.
We have also computed the energy per particles at rho=0.175 and T=0.05, 
and compared with the results in Table VII of Ref\cite{caillol1995}: our 
results is $\langle U/N k_BT \rangle =-12.38 \pm 0.01$  in perfect agreeement 
with their results of $U/Nk_BT=-12.38$. Moreover, 
we computed the energy per particle 
at lower densities where the system is in a vapour phase, 
at $\rho^{*}=0.002$ and $T^{*}=0.05$\cite{caillol1995}, and found 
$\langle  U/Nk_BT \rangle=-10.20 \pm 0.05$, 
in good agreement with that computed for the 
same state point by Caillol and Weis\cite{caillol1995} ($U/Nk_BT=-10.15$).

It is well known that the RPM forms clusters in the subcritical vapour 
phase. In order to identify the clusters, we use Gillan's definition, 
according to which two particles belong to the same cluster if they are 
separated by a distance shorter than a given cut-off $\lambda$ 
\cite{gillan1983}. In this way, we detect the total number of isolated 
ions, the total number of associated ions, and the total number of 
clusters of a given size. In what follows, and unless stated otherwise, we 
will study the cluster formation when $\lambda = 2\sigma$.

\section{Techniques to equilibrate the RPM at low temperatures and 
densities}

In order to study the RPM at low temperatures and densities, {\it ad-hoc} 
simulation methods have been employed to overcome the problem of slow 
convergence towards equilibrium. The main obstacle to simulate the RPM in 
the low-temperature region is the strong binding effect of oppositely 
charged ions at short distances, as the thermal energy available to drive 
two oppositely charged particles away from each other is much less than 
the attractive Coulomb energy, i.e., $T^{*} \ll 1$. Therefore, in order to 
reach equilibrium, the system would have to be simulated for a 
prohibitively long time. In our simulations, this equilibration problem is 
going to be even more pronounced, since we aim to study very dilute 
systems where isolated ions are so far apart from each other that they 
spend most of the time freely diffusing in the empty space. Once they 
finally find an oppositely charged ion, they strongly bind to it forming a 
neutral dimer (or a higher cluster) that rarely breaks. As a consequence, 
the computational time needed to equilibrate the system can be 
astronomically long. To improve the equilibration time in our $NVT$ MC 
scheme, we have adopted two established MC moves and implemented a new 
one:
\begin{itemize}

\item small and large particle displacements;

\item small and large cluster displacements;

\item formation and breakage of clusters.

\end{itemize}

\subsection{Small and large particle displacements}

The first move we select is the standard single-particle displacement, 
where the $x$, $y$, and $z$ coordinates of a randomly selected particle 
are each displaced by a small amount $\delta$ chosen randomly from the 
interval $\{-\delta_{\rm max},\delta_{\rm max}\}$. The move is accepted 
with the standard Metropolis probability $\min{(1,e^{-\beta [ u(n) - 
u(o)]})}$, where $u(o)$ and $u(n)$ are the energies of the particle before 
and after the trial move, respectively, and $\beta=1/k_BT$. According to 
normal practice, $\delta_{\rm max}$ can be adjusted to give some desired 
acceptance rate for the move over the course of the simulation. In 
principle, such moves should allow each particle to diffuse as a free ion, 
and join or leave a cluster. However, when the density is very low (and 
the simulation box is very large), short single-particle displacements are 
not sufficient to sample the phase space properly. Thus, we also randomly 
attempt displacements where $\delta_{\rm max}=L/2$. These occasional large 
displacements are intended to accelerate cluster formation (if 
thermodynamically favourable) and to allow the system to explore more 
significant regions of phase space within the simulation timescale.

\subsection{Small and large cluster displacements}

Single-particle MC moves are not enough to equilibrate highly clustered 
systems, and so we also implement a cluster move similar to that proposed 
by Orkoulas and Panagiotoupoulos \cite{orkoulas1994}: we first identify 
all of the clusters in the system, then choose a cluster at random and 
select displacements from either a small or a large interval, as in the 
single-particle moves. Cluster moves that result in the merging of two or 
more clusters have to be treated extremely carefully in order to respect 
detailed balance; the reverse move has to be attempted with equal 
probability as the forward move. Here we take a simple solution, and 
simply reject all cluster moves that lead to the merging of clusters 
\cite{orkoulas1994,caillol1995}. In this way, the instantaneous cluster 
distribution is left intact, and detailed balance cannot be violated. With 
this simple approach, the cluster move is accepted with the normal 
Metropolis probability $\min{(1,e^{-\beta [U(n) - U(o)]})}$, where $U(o)$ 
and $U(n)$ are the energies of the system before and after the trial move, 
respectively. Of course, this move does not lead to the formation or 
breakage of clusters; a specific move to effect these transformations is 
detailed next.

\subsection{Novel move for the formation and breakage of clusters}

The last attempt we make to improve the equilibration of the system is to 
introduce a novel move that offers the opportunity of forming and breaking 
clusters. This `cluster formation/breakage' (CFB) MC move is designed to 
respect detailed balance, and is implemented as follows:
\begin{enumerate}

\item we choose a particle at random (particle 1), without knowing {\it a 
priori} whether it belongs to a cluster;

\item we identify all of its neighbours within a cut-off distance 
$\Delta$, which can be tuned to give optimal performance, as described 
below;

\item we choose a neighbour at random (particle 2), irrespective of its 
charge, and store its separation from particle 1, $r_{12}(o)$;

\item we then move particle 2 to a new separation from particle 1, 
$r_{12}(n)$, chosen randomly and uniformly from the interval $\sigma \leq 
r_{12}(n) \leq \Delta$, and with a random orientation of the corresponding 
separation vector;

\item we accept the move with a probability 
$\min{(1,[r_{12}(n)/r_{12}(o)]^2 e^{-\beta [U(n) - U(o)]})}$, where $U(o)$ 
and $U(n)$ are the energies of the system before and after the trial move, 
respectively.

\end{enumerate}
This CFB move respects detailed balance, and does not add any bias towards 
the formation or breakage of a cluster, since clusters can form and break, 
and isolated ions can simply be displaced (see Appendix).

In the current work, the target acceptance rate for the single-particle 
and cluster moves is approximately 40\%, while the acceptance rate for the 
CFB moves varies between a few percent (at high density) and 40\% (at very 
low density).

\section{Results}

\begin{figure}[ht]
\begin{center}
\includegraphics[height=6.5cm,clip=true]{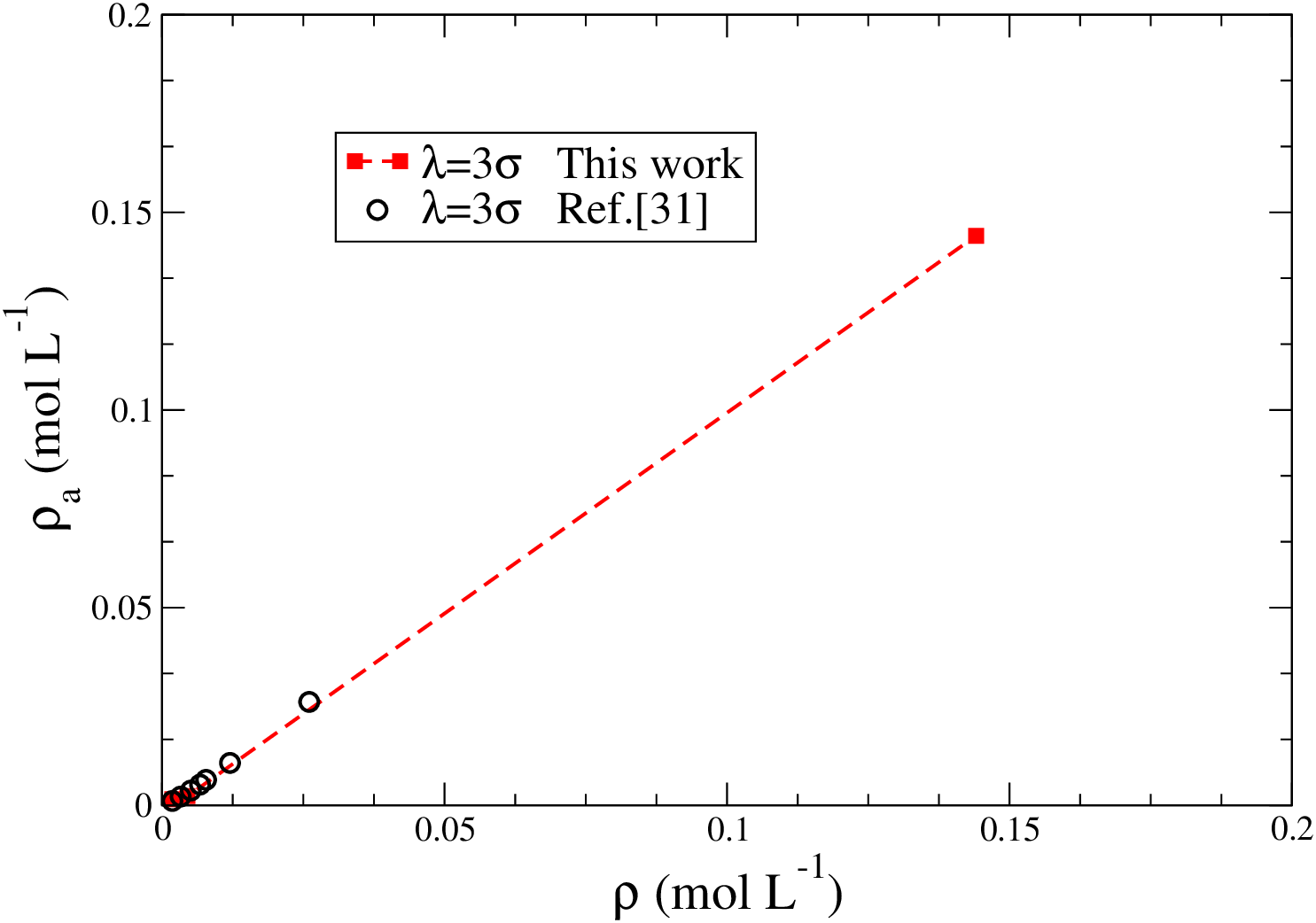}
\caption{Concentration of associated ions $\rho_{a}$ against the total ion 
concentration $\rho$. The open circles are results from reference 
\cite{erratum} and the red squares are our results. The cut-offs chosen to 
identify the clusters are indicated in the legend.} \label{complowen}
\end{center}
\end{figure}

We start by comparing our simulation results (using single-particle, 
cluster, and CFB moves) with those obtained by Allahyarov {\it et al.} 
\cite{erratum} under the conditions $T=300~\mbox{K}$, $\epsilon=8$, $q=e$, 
and $\sigma=10~\mbox{\AA}$, corresponding to a reduced temperature $T^{*} 
\simeq 0.14$. The molar concentrations of salt lie in the range 
$10^{-4}$-$10^{-2}~\mbox{mol}~\mbox{L}^{-1}$. Allahyarov {\it et al.} 
counted ``the number of oppositely charged pairs which are closer than 
$3\sigma$", whereas we consider associated ions belonging to clusters with 
two or more ions (with the same cut-off). Figure \ref{complowen} shows the 
total concentration of associated ions $\rho_{a}$ as a function of the 
total ion concentration $\rho$. Data from figure 1 of reference 
\cite{erratum} are included after multiplying the salt concentration and 
associated ion-pair concentration by two; in keeping with these data, we 
quote the concentrations in units of $\mbox{mol}~\mbox{L}^{-1}$, assuming 
a particle diameter $\sigma=10~\mbox{\AA}$. All we want to emphasise here 
is that we get good agreement with the established results for 
concentrations in the range $10^{-4}$-$10^{-2}~\mbox{mol}~\mbox{L}^{-1}$.

After having confirmed that the algorithm is working in the density regime 
that has already been studied, we move to the central aim of this 
manuscript, i.e., equilibrating the RPM at extremely low density. The 
lower the temperature and the density, the longer it takes to equilibrate 
the system. Thus, a good test for the algorithm is to equilibrate the 
system at the lowest density and lowest temperature of interest, i.e., 
$T^{*}=0.04$ and $\rho^{*} = 9.03 \times 10^{-11}$. From preliminary 
tests, it appears that at the same temperature and higher densities of 
around $\rho^{*} = 10^{-6}$ the system equilibrates in a reasonably short 
time.

For clarity, we define three different MC cycles: Monte Carlo cycle 0 
(MC0) consists of $N$ moves, 90\% of which are small displacements of 
single particles, 3\% are large displacements of single particles, 3\% are 
are small displacements of randomly chosen clusters, and 4\% are large 
displacements of randomly chosen clusters; Monte Carlo cycle 1 (MC1) 
consists of $N$ moves, 90\% of which are small displacements of single 
particles, 3\% are large displacements of single particles, 3\% are small 
displacements of randomly chosen clusters, 2\% are large displacements of 
randomly chosen clusters, and 2\% are CFB moves; Monte Carlo cycle 2 (MC2) 
consists of $N$ moves, 70\% of which are small displacements of single 
particles, 10\% are large displacements of single particles, 5\% are small 
displacements of randomly chosen clusters, 5\% are large displacements of 
randomly chosen clusters, and 10\% are CFB moves. In all cases, we define 
a cluster according to $\lambda=2\sigma$; in MC1 we choose $\Delta=L/2$, 
whereas in MC2 we consider different values of $\Delta$.

\begin{figure}[ht]
\begin{center}
\includegraphics[height=6cm,clip=true]{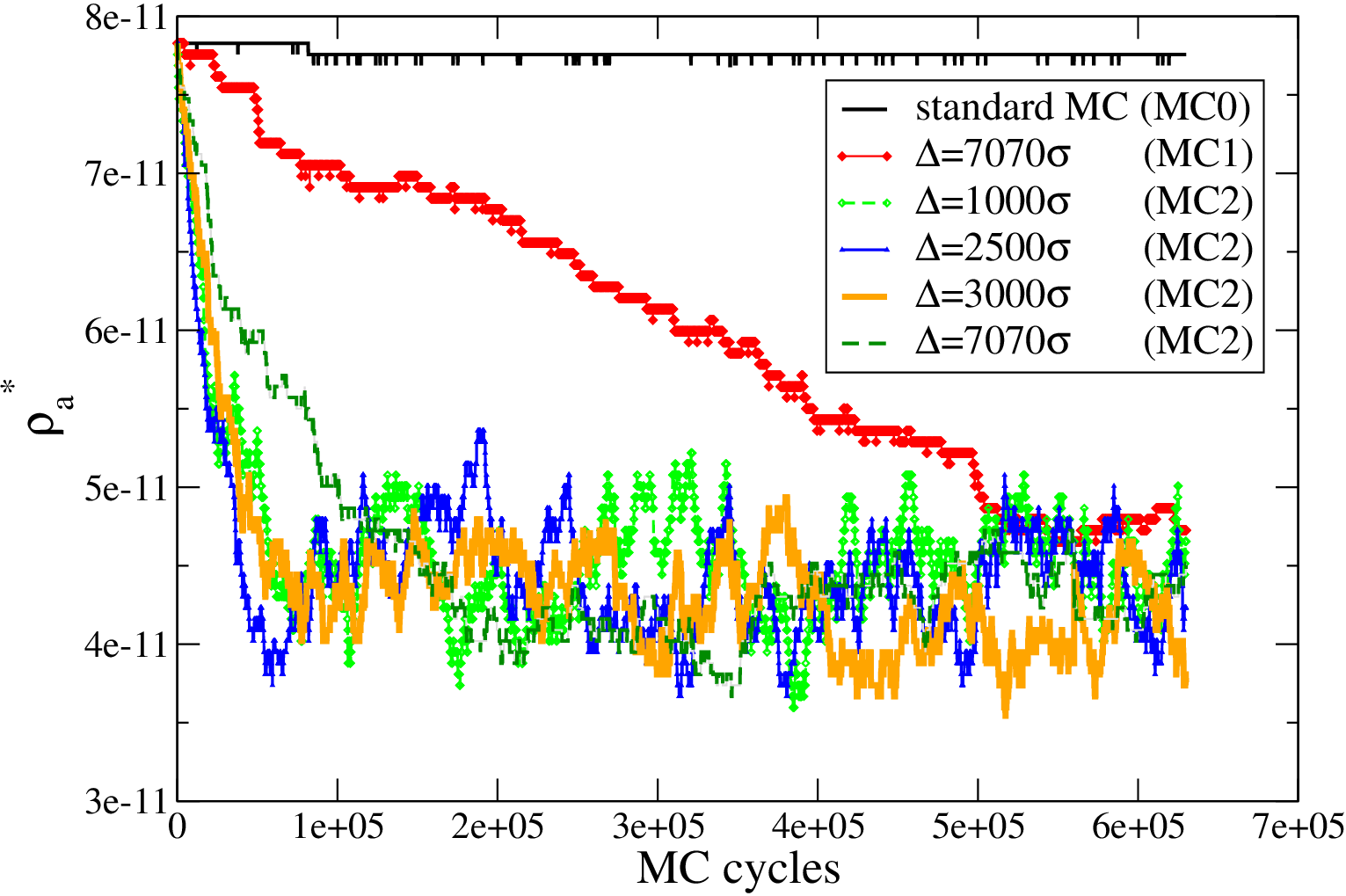}
\caption{Reduced density of associated ions versus Monte Carlo cycle for 
the MC0, MC1 and MC2 protocols and the same initial configuration at 
$T^{*}=0.04$ and $\rho^{*}= 9.03  \times 10^{-11}$. 
The legend indicates the cut-off 
$\Delta$ chosen for CFB.\label{comparedbf}}
\end{center}
\end{figure}

In figure \ref{comparedbf} we show the reduced density of associated ions, 
$\rho_{a}^{*}$, versus MC cycle for simulations run according to the MC0, 
MC1, and MC2 protocols, and starting from the same initial configuration. 
MC0 shows almost no structural evolution on the simulation timescale, and 
hence is entirely inadequate for simulations at low temperature and 
density. This is caused by the incredibly long distances an ion should 
cover in order to find another ion (difficult cluster formation), and at 
the same time by the low probability of thermally activated dissociation 
of ion pairs at very low temperature (difficult cluster breakage). It is 
evident that all of the MC2 runs and the MC1 run equilibrate to the same 
structure, within our simulated time scale. 
Moreover, the equilibration times are quite different: MC2 (with 10\% CFB 
moves) equilibrates faster than MC1 (with 2\% CFB moves).

Next, we select the MC2 Monte Carlo scheme and equilibrate the system 
using different values of $\Delta$, to show that the final equilibrium 
state does not depend on the choice of $\Delta$, but that its 
equilibration rate does. To this end, we use different values of $\Delta$, 
ranging from 1000$\sigma$ up to half of the box length $L/2$ 
(7070$\sigma$), and plot the reduced density of associated ions versus MC 
cycle. Figure \ref{comparedbf} shows that all of the chosen values of 
$\Delta$ lead to the same equilibrium density of associated ions. 
Strikingly, the equilibration rate decreases with increasing $\Delta$. 
Choosing a small value for $\Delta$ allows for a faster equilibration; 
however, $\Delta$ cannot be too small compared to the mean separation of 
clusters, as it will lead again to inefficient sampling, not allowing 
clusters to merge or break. Therefore, the optimal value of $\Delta$ 
should decrease with increasing density.

\begin{figure}[ht]
\begin{center}
\includegraphics[height=6.5cm,clip=true]{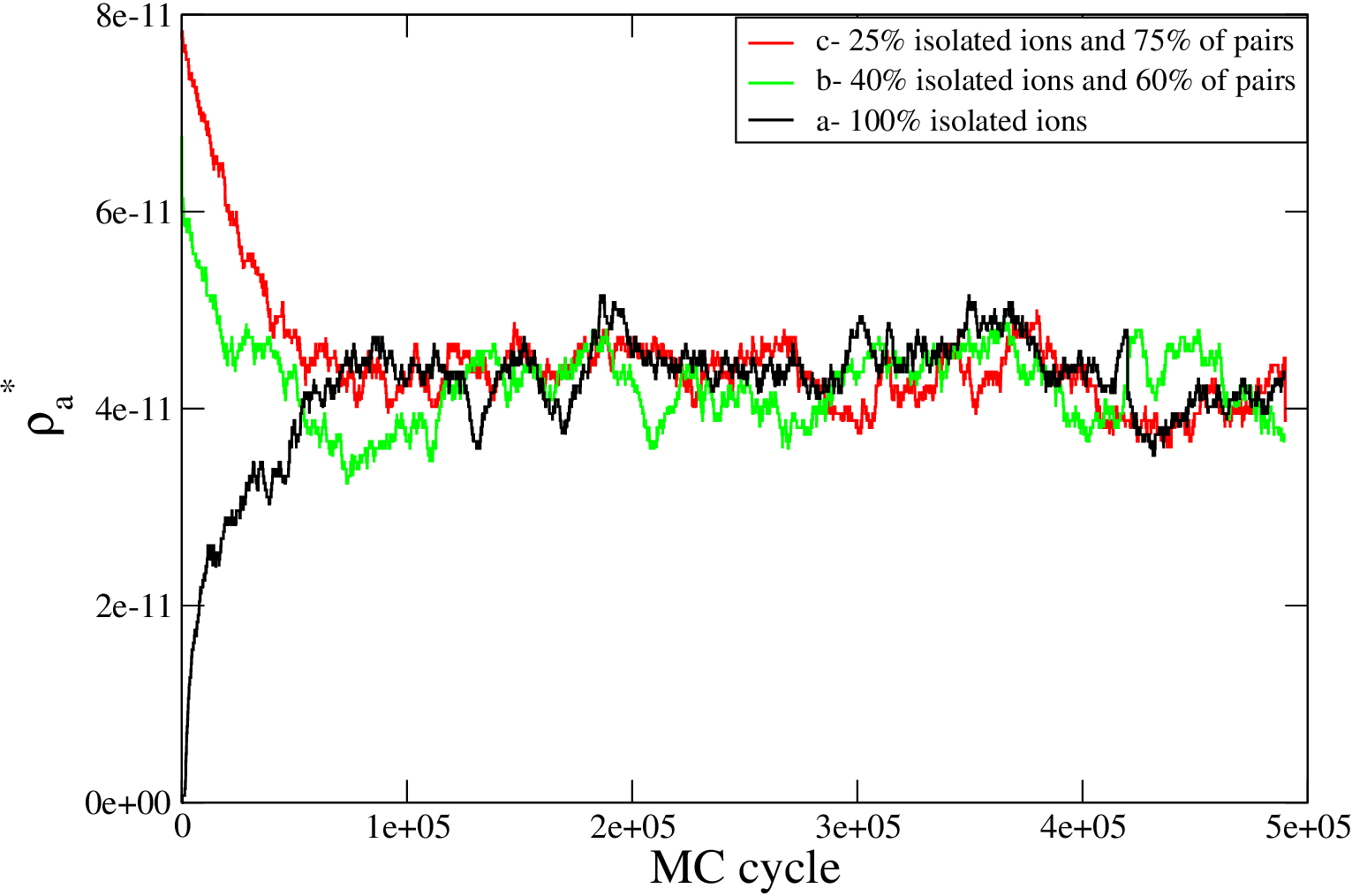}
\caption{Reduced density of associated ions versus Monte Carlo cycle in 
simulations started from configurations (a), (b), and (c) (see text), at 
$T^{*}=0.04$ and $\rho^{*}= 9.03  \times 10^{-11}$.\label{compare54all}}
\end{center}
\end{figure}

We now demonstrate that the convergence of the algorithm does not depend 
on the initial configuration chosen, and that the system is quasi-ergodic 
on the simulation time scale. To this end, we set $\Delta=1000 \sigma$ and 
compute the density of associated ions in simulations starting from three 
completely different initial configurations: (a) a configuration 
containing only isolated ions; (b)  a configuration containing $40\%$ 
isolated ions and 60\% ions in pairs; and (c) a configuration containing 
25\% isolated ions and 75\% ions in pairs. Figure \ref{compare54all} shows 
that convergence is achieved irrespective of the initial configuration. It 
is also encouraging that the algorithm allows for significant fluctuations 
in the number of associated ions, which indicates that there is a dynamic 
equilibrium involving the formation and breakage of clusters.

\section{Conclusions}

In this manuscript we have presented a numerical method for speeding-up 
computer simulations of the restricted primitive model at low temperatures 
(around $T_{c}$) and very low reduced densities (down to $10^{-10}$). Our 
method involves the combination of conventional single-particle and 
cluster moves with a novel `cluster formation/breakage' move, designed 
specifically to equilibrate the system in a reasonable time, even at such 
extreme thermodynamic conditions. The suggested Monte Carlo scheme is 
straightforward to implement: after having set the value of the maximum 
neighbour distance $\Delta$ the method is inherently efficient, in that 
the system quickly converges to its equilibrium state. This method might 
also be applied to the equilibration of other systems that form aggregates 
at low concentrations. We should mention that we are aware of other 
techniques that might prove useful to equilibrate very low concentration 
systems, such as the `geometric cluster algorithm' by Liu and Luijten 
\cite{luijten2004}, demonstrated to speed up simulations of complex fluids 
near criticality and/or with differently sized components, and a novel 
cluster move by Almarza \cite{noe2009,noe2009.1}. As far as we are aware, 
our results extend to far lower concentrations than in any previous 
studies on the vapour phase of the restricted primitive model. The 
algorithm presented here allows for a comprehensive study of the vapour 
phase around the critical temperature and at reduced densities down to 
$10^{-10}$: such low densities seem to be relevant for experiments on 
low-concentration solutions of ions in low-dielectric organic solvents. A 
detailed report of our investigations is in preparation.

\section*{Acknowledgements}

C.\ V.\ thanks A.\ Cuetos, D.\ Frenkel, D.\ Marenduzzo and E.\ Sanz for 
valuable discussions and suggestions. This work was, at an early stage, 
financially supported by an NWO-VICI grant. C.V. is supported by an 
Individual Inter-European Marie Curie Fellowship. Computer resources were 
provided by AMOLF (The Netherlands).

\section*{Appendix: Acceptance move of the cluster formation/breakage 
move} \setcounter{equation}{0}

Below we derive the acceptance rule for the CFB move, and show that it 
satisfies the detailed balance condition. Detailed balance requires that
\begin{equation}
\label{db1}
p(o) \pi(o \rightarrow n) = p(n) \pi(n \rightarrow o)
\end{equation}
where $p(o)$ is the probability that the system is initially in the old 
configuration $o$, $\pi(o \rightarrow n)$ is the transition probability 
from the old to the new configuration $n$, $p(n)$ is the probability the 
system is initially in the new configuration, and $\pi(n \rightarrow o)$ 
is the transition probability from the new to the old configuration. Each 
transition probability in equation (\ref{db1}) can be expressed as the 
product of two terms:
\begin{equation}
\label{db2}
\pi(o \rightarrow n) = 
\alpha (o \rightarrow n) \times acc(o \rightarrow n).
\end{equation}
$\alpha(o \rightarrow n)$ is the probability of generating a new 
configuration $n$ starting from $o$, and $acc(o \rightarrow n)$ is the 
probability of accepting the move. A similar equation holds for $\pi(n 
\rightarrow o)$. In our simulations, the old configuration is defined by 
choosing two particles (1 and 2) at random, and computing their relative 
distance $r_{12}(o)$, and the total energy of the system $U(o)$; the new 
configuration is generated by displacing particle 2 with respect to 
particle 1, and computing their new relative distance $r_{12}(n)$, and the 
new total energy of the system $U(n)$. $r_{12}(n)$ is generated uniformly 
on the interval $\{\sigma,\Delta\}$, and hence $\alpha (o \rightarrow n) = 
\alpha (n \rightarrow o)$. The Boltzmann probability goes like $p \propto 
r_{12}^{2} e^{-\beta U}$. Combining equations (\ref{db1}) and (\ref{db2}) 
gives
\begin{equation}
\label{db3}
\frac{acc(o \rightarrow n)}{acc(n \rightarrow o)} =
\frac{\alpha(n \rightarrow o)}{\alpha(o \rightarrow n)}
\frac{p(n)}{p(o)} =
\left[ \frac{r_{12}(n)}{r_{12}(o)} \right]^{2}
e^{-\beta[U(n)-U(o)]}.
\end{equation}
To conclude, we implement a Metropolis sampling scheme using an acceptance 
probability for a move from $o$ to $n$ of
\begin{equation}
\label{db5}
acc(o \rightarrow n) = 
\min \left( 1, \left[ \frac{r_{12}(n)}{r_{12}(o)} \right]^{2}
e^{-\beta[U(n)-U(o)]} \right).
\end{equation}

\section*{References}

\bibliography{bjerrum.10}

\end{document}